\documentclass[aps,prb,showpacs,twocolumn]{revtex4-1}
\usepackage{graphicx}
\usepackage{epsfig}
\usepackage{amssymb, amsmath}
\usepackage{float}
\usepackage{dcolumn}
\usepackage{bm}

\begin{document}

\title{\bf Current carried by evanescent modes and possible device application}
\author{ Sreemoyee Mukherjee$^1$, P. Singha Deo$^1$, and A. M. Jayannavar$^2$}
\affiliation{$^1$Unit for Nanoscience and Technology, S.N.Bose National Centre For Basic Sciences\\Block-JD, 
Sector-III, Salt Lake, Kolkata-700098, India\\
$^2$ Institute of Physics, Sachivalaya Marg, Bhubaneswar 751005, India.}
\date{\today}
\begin{abstract}
Quantum tunneling of an electron through a classically forbidden
regime has no classical analogue and several aspects of it are
still not well understood. In this work we analyze electronic currents 
under the barrier. For this we consider
a multichannel Aharonov-Bohm ring and develop a correct
formalism to calculate the currents inside the ring when the
states are evanescent. We also show unlike other proposed
quantum devices that such currents and associated conductance are not very sensitive to
changes in material parameters and thus the 
system can be used to build stable devices that work on
magnetic and transport properties. We also study the current
magnification property of the ring in presence of both propagating
and evanescent states.
\end{abstract}
\maketitle
\section{\bf Introduction}
In the last few decades, there has been a major interest in the
area of mesoscopic systems \cite{els}. Most of the experimental and
theoretical studies of these systems have involved transport
measurements, usually of the resistance as a function of temperature
or magnetic field. However, there is increasing interest in other
non-transport properties, such as magnetization \cite{levy64, chand, maily}. 
This has prompted a lot of experimental works, including the study of arrays
of mesoscopic systems \cite{levy64,chand}, and the use of very small Hall sensors and
microsquid magnetometers \cite{geim}. These studies give us an unique opportunity
to understand and explore manifestations of quantum mechanics like 
persistent currents \cite{96A}, circulating currents \cite{jay1} etc.

Mesoscopic structures have led to the possibility of new semiconductor devices. 
These quantum devices rely on quantum effects for their operation 
based on quantum interferometric principles, and are quantum analogs of well-known 
optical and microwave devices \cite{datta}. 
Several potential magnetic devices have been proposed 
\cite{pop}, viz., Hall devices, magnetoresistors, 
inductive proximity and distance sensors, fluxgate sensors, other 
magnetic sensors such as magnetodiodes, magFETs, magnetotransistors 
and carrier domain magnetometers. Some commonly seen devices that
work in classical regime and use magnetic properties are magnetic tape
used for data storage, magnetic card reader, keycard lock etc. 
The conventional devices operate in a classical diffusive regime 
and are not very sensitive to variations in material parameters such as dimensions or 
the presence of small impurities or non-uniformity in size and shape. These devices 
operate by controlling the carrier density of quasi-particles.
So far no quantum devices in mesoscopic or nanoscopic length 
have been achieved practically because quantum devices have an
inherent shortcoming. Proposed quantum devices are not very robust in the sense 
that the operational characteristics depend very sensitively on material parameters, 
impurity configuration, shape and size of sample, temperature and 
Fermi energy \cite{land}. For example, incorporation of a single 
impurity in the mesoscopic device can change, 
non-trivially, the interference of partial electron waves propagating through 
the sample, and hence the electron transmission (operational characteristics) 
across the sample \cite{bcg}. In such devices the actual problem of control 
and reproducibility of operating thresholds become highly nontrivial.
These devices can be exploited if we achieve the 
technology that can reduce or control the phase fluctuations \cite{deo} to a small fraction 
of $ {2\pi} $. Devices in which electrons carry current without being 
scattered either elastically or inelastically (ballistic devices) promise 
to be much faster and will consume less power than the conventional devices. 
It should also be noted that quantum devices can 
exhibit multifunctional property (e.g., single stage frequency multiplier) wherein 
the functions of an entire circuit within a single element can be performed 
\cite{sub}. They can also lead to down sizing of electronic devices.

The conductance of a quantum ring is an oscillating function of 
magnetic flux through the ring 
implying we can use flux to drive the system from a conducting state to an
insulating state which can be identified as 1 and 0 of a switch as will be exemplified.
Such a switch will therefore be working entirely on quantum interference principles
which is a new idea in electronic application. Several potential switching devices
have been proposed, wherein one controls the relative phase difference
between different interfereing paths (in semiconducting loop structure) 
by applying electrostatic or magnetic fields \cite{dat1, dat2, dat3}. The
possibility of achieving such transistor action in T-shaped structure by varying
the effective length of the vertical open ended lead has also been explored
\cite{sol}. All these proposed devices has not been practical so far because
of the inherent shortcoming of quantum devices discussed above.

The magnetization of the ring 
can be used to store information just like a spatial array of capacitors store
information in present day computers or spatially varying magnetic field
in magnetic tapes. 
There can be an array of many quantum rings of different sizes. These can cause a magnetization which has complex spatial variation.
A lot of work has been done in one dimensional 
quantum rings \cite{deo, mol, pet, pet1, kal, jos, ben2,pas}. 
However, the experimental rings are always two dimensional or three 
dimensional. Such systems have not received much theoretical attention 
because multichannel junctions are very difficult to treat theoretically. 
Open rings where particle exchange can occur and temperature can be defined 
is more general as closed
ring properties can be seen as a special case \cite{akk}. This will be
more realistic to study. Besides, open rings can exhibit some
novel properties like current magnification or evanescent modes that
have no analogs in closed systems. Current magnification is the phenomenon
where due to quantum interference large current can circulate in a loop.
Circulating currents (or heat current magnification) can also arise in
classical coupled oscillator loops \cite{pre}. Spin current magnification can 
also exhibit magnification effect \cite{choi}. 
Earlier models either do not account for channel mixing or do not allow 
the inclusion of evanescent modes. In our present work we account
for both of these allowing us to study evanescent states in realistic
quasi one-dimensional systems. 
We show that magnetization due to such evanescent states 
have very interesting manifestation of quantum effects. We also argue 
that such effects can be used to build stable devices that uses 
quantum interference effects. We also study current magnification effect.

\section{\bf Theoretical Analysis}
\begin{figure}[h!]
\centering
\includegraphics[height=6cm,width=9cm]{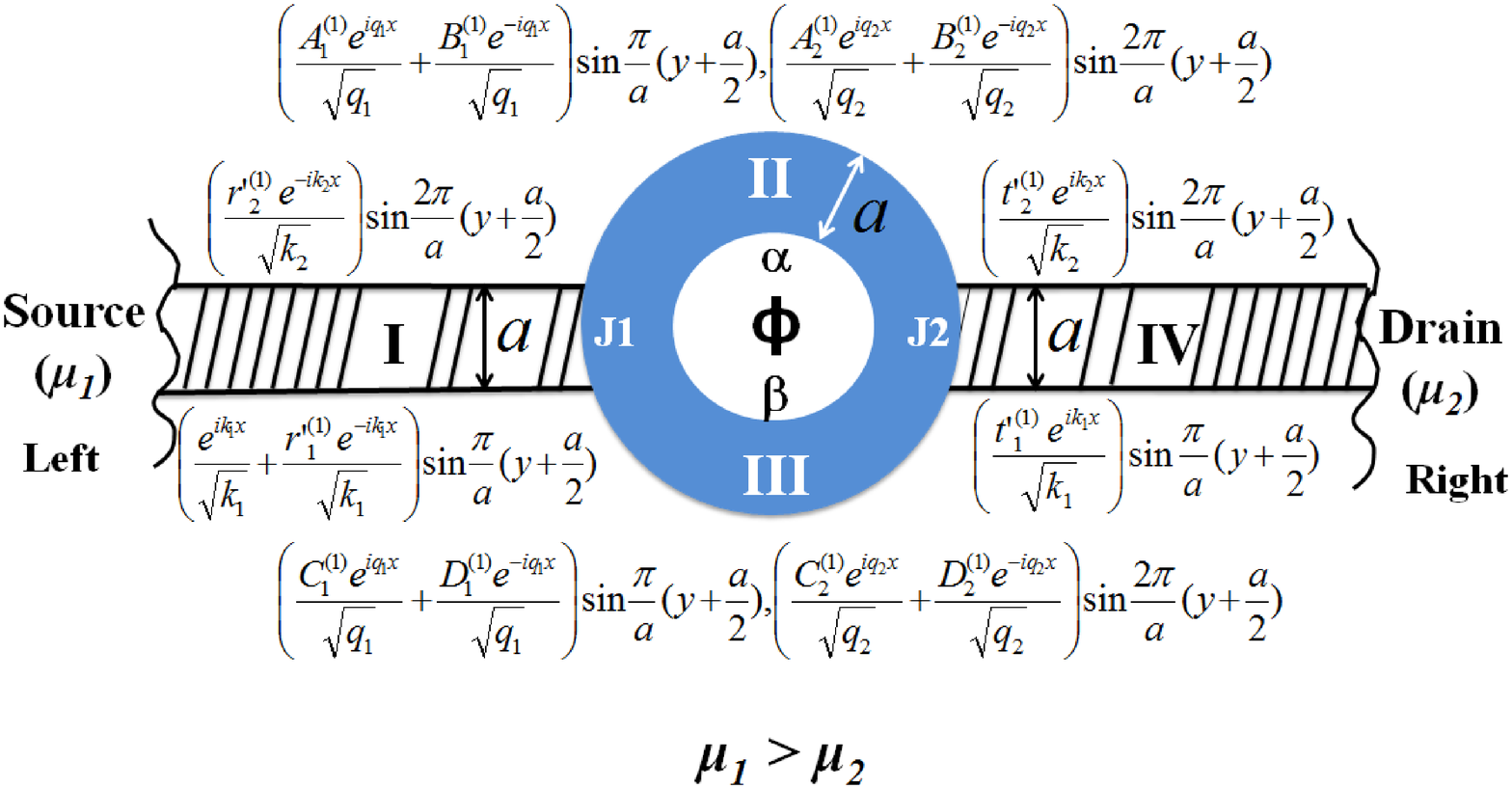}
\caption{\small A finite thickness quantum ring of width $ a $ made 
up of normal metal or semiconductor is indicated by the grey region. 
On either sides the quantum ring is attached with quantum wires (stripped
region) made up of normal metal or semiconductor. On the left of the 
above system there is the source reservoir whose chemical potential 
is $\mu_{1}$ and on the right there is the drain reservoir whose 
chemical potential is $\mu_{2}$.  A potential difference $(\mu_{1}-\mu_{2})$ 
drives transport currents through the system. The wave functions of 
the electron in different regions is 
shown in the figure at their respective places. Different regions 
are marked as I, II, III and IV. The ring is pierced by an 
Aharonov-Bohm (A-B) flux $ \phi $. $ \alpha $ is the A-B
phase an electron picks up in region II and $ \beta $ is that in 
region III. J1  is the junction where the regions I, II and III 
meet and J2  is the junction where the regions II, III and IV meet.}
\label{fig2}
\end{figure}
Fig. \ref{fig2} represents the schematic diagram of a quantum circuit 
consisting of a ring geometry, made of finite thickness quantum wires. Such quantum wires act like a 
waveguide \cite{prb50, prb58}. Electronic transport in such 
systems can be well described by an effective mass theory 
\cite{datta}. Incident electrons coming from the source reservoir 
on the left (say), gets scattered by the ring. Division of wave 
front occurs at junction J1; a partial wave propagates along 
the upper arm of the ring and another partial wave propagates along the 
lower arm of the ring. These two partial waves recombine and give a 
transmittance that bears the signature of interference between 
the two partial waves along the two arms of the ring. This 
interference can be modified by an Aharonov-Bohm (A-B) flux through 
the center of the ring. The description of the figure is given 
in further detail in the figure caption. 

The Schr$\ddot{o}$dinger equation for a quasi one dimensional wire is (the third 
degree of freedom, i.e., $z$-direction, is usually frozen by creating a strong quantization \cite{datta})  
\begin{equation}
-{{{\hbar}^2} \over 2m^*}({{\partial}^2{\psi} \over {{\partial}x^2}}+
{{\partial}^2{\psi} \over {{\partial}y^2}}) + V(x,y){\psi}(x,y) = E{\psi}(x,y)
\label{eq1}
\end{equation} 
Here the $ x $ coordinate is along the wire, $y$ coordinate is perpendicular to it, $ m^* $ 
is the electron effective mass and $E$ is the electron energy. In regions I and IV (see Fig. \ref{fig2}) 
we have only the confinement potential. That is $$ V(x,y) =  V(y) .$$ Whereas in regions II and 
III apart from the confinement potential we take a constant potential $ V_0 $ that can be used 
to excite evanescent modes inside the ring. That is $$ V(x,y) = V(y) + V_0 $$ Without any loss
 of generality we take $ V(y) $ to be an infinite square well potential
of width $a$. That is\\
\[V(y) = \left\{ \begin{array}{ll}
         0 & \mbox{ if $-a/2 < y < a/2$};\\
         \infty & \mbox{ if $|y|\geq a/2$}.\end{array} \right. \]
The wave functions in the ring can be obtained by solving Eq. (\ref{eq1}) 
where we assume the ring to be so large compared to the de Broglie wave 
length that its curvature can be neglected \cite{she}. The length of the 
ring is $ L= l_{U} + l_{L} $, where $ l_{U} $ is the length of the upper 
arm and $ l_{L} $ is the length of the lower arm. The magnetic field appears 
just as a phase of $ {\psi}(x,y) $ and will be accounted for while applying 
the boundary conditions \cite{che}. In regions I and IV Eq. (\ref{eq1}) can 
be separated as
\begin{equation} 
\psi(x,y) = \phi(x)\xi(y) 
\label{eq3}
\end{equation}
to give
\begin{equation}
-{{{\hbar}^2} \over 2m^*}{{{\partial}^2{\phi}(x)}\over {{\partial}x^2}} = {{{\hbar^2}k^2}\over 2m^*}{\phi}(x) 
\label{eq4} 
\end{equation}
and
\begin{equation}
-{{{\hbar}^2} \over 2m^*}{{{\partial}^2{\xi}(y)}\over {{\partial}y^2}}+V(y){\xi}(y) = \varepsilon{\xi}(y) 
\label{eq5}
\end{equation}
Since $ V(y)$ is an infinite square well potential of width $ a $, Eq. (\ref{eq5}) gives 
\begin{equation}
{\xi}_n(y)={\sin}{{n\pi}\over a}({a\over 2} +y)
\label{eq6}
\end{equation} and 
\begin{equation}
\varepsilon_n = {{n^2{\pi}^2{\hbar}^2}\over{2m^*a^2}}
\label{eq7} 
\end{equation}
Eq. (\ref{eq4}) has solution of the form 
$$\phi_n(x) = e^{{\pm}ik_nx}$$ with 
\begin{equation}
k_n=\sqrt{\frac{2m^*E}{\hbar^2}-\frac{n^2\pi^2}{a^2}} 
\label{eq8}
\end{equation}
or 
\begin{equation}
E = \varepsilon_n + {{{\hbar^2}k_n^2}\over 2m^*}
\label{eq9}
\end{equation}
So wave function in region I for electrons incident along channel 1 on the 
left and electrons are also reflected back to channel 1 on the left, can be written as
\begin{equation}
{\psi(I)}^{(1)}_1 = ({e^{ik_1x}\over\sqrt k_1}+{r'^{(1)}_{1}e^{-ik_1x}\over\sqrt k_1})\sin{{\pi}\over a}(y+{a\over 2}) 
\label{eq10}
\end{equation}
Wave function in region I for electrons incident along channel 1 on the left 
and electrons are reflected back to channel 2 on the left, can be written as
\begin{equation}
{\psi(I)}^{(1)}_2 = ({r'^{(1)}_{2}e^{-ik_2x}\over\sqrt k_2})\sin{2{\pi}\over a}(y+{a\over 2}) 
\label{eq11}
\end{equation}
Wave function in region IV for electrons incident along 
channel 1 on the left and electrons are transmitted to channel 1 on the right can be written as
\begin{equation}
{\psi(IV)}^{(1)}_1 = ({t'^{(1)}_{1}e^{ik_1x}\over\sqrt k_1})\sin{{\pi}\over a}(y+{a\over 2})
\label{eq12}
\end{equation}
Wave function in region IV for electrons incident along 
channel 1 on the left and electrons are transmitted to channel 2 on the right can be written as
\begin{equation}
{\psi(IV)}^{(1)}_2 = ({t'^{(1)}_{2}e^{ik_2x}\over\sqrt k_2})\sin{2{\pi}\over a}(y+{a\over 2})
\label{eq13}
\end{equation}
From Eq. (\ref{eq8}), in the first mode
\begin{equation}
k_1=\sqrt{\frac{2m^*E}{\hbar^2}-\frac{\pi^2}{a^2}}
\label{eq14}
\end{equation}
is the propagating wave vector and in the second mode 
\begin{equation}
k_2=\sqrt{\frac{2m^*E}{\hbar^2}-\frac{4 \pi^2}{a^2}}
\label{eq15}
\end{equation}
is the propagating wave vector. For 
\begin{equation}
\frac{4 \pi^2}{a^2} < E < \frac{9 \pi^2}{a^2}
\label{eq16}
\end{equation}
both $ k_1 $ and $ k_2 $ are real as can be seen from Eq. (\ref{eq14}) and Eq. (\ref{eq15}). 
In this energy range $ k_n $ for $ n>2 $ are imaginary as can be seen from Eq. (\ref{eq8}) implying that 
there are two propagating channels. In the leads we can not have evanescent modes \cite{wee,wha}. \\

Now for the regions II and III the potential is $ V(x,y)= V(y) + V_0 $. 
In these regimes using $q_n$ as wave vector, the energy can be written 
similarly like region I and IV as follows
\begin{equation}
E - V_0 = {{{\hbar}^2q_n^2}\over 2m^*} + {{n^2{\pi}^2{\hbar}^2}\over{2m^*a^2}} 
\label{eq17}
\end{equation}
or
$$ q_n = \sqrt{{2m^*(E - V_0)\over {\hbar}^2} - {n^2{\pi}^2\over {a^2}}} $$
Hence, in the first mode
\begin{equation}
q_1 = \sqrt{{2m^*(E - V_0)\over {\hbar}^2} - {{\pi}^2\over {a^2}}}
\label{eq18}
\end{equation}
is the wave vector and in the second mode 
\begin{equation}
q_2 = \sqrt{{2m^*(E - V_0)\over{\hbar}^2} - {{4\pi^2}\over {a^2}}}
\label{eq19}
\end{equation}
is the wave vector.
In regions II and III, three different situations can arise depending on the choice
of energy, $E$, potential $V_0$ and $n$. Wave functions in these regimes for 
$ E > V_0 + {n^2 \pi^2 \hbar^2\over 2m^{*}a^2} $ can be written similarly as in 
Eq. (\ref{eq10}) - Eq. (\ref{eq13}). 
\begin{equation}
\psi(II)^{(1)}_1 = ({A^{(1)}_{1}e^{iq_1x}\over\sqrt q_1}+{B^{(1)}_{1}e^{-iq_1x}\over\sqrt q_1})\sin{{\pi}\over a}(y+{a\over 2})
\label{eq20}
\end{equation}
\begin{equation}
\psi(II)^{(1)}_2 = ({A^{(1)}_{2}e^{iq_2x}\over\sqrt q_2}+{B^{(1)}_{2}e^{-iq_2x}\over\sqrt q_2})\sin{2{\pi}\over a}(y+{a\over 2})
\label{eq21}
\end{equation}
\begin{equation}
\psi(III)^{(1)}_1 = ({C^{(1)}_{1}e^{iq_1x}\over\sqrt q_1}+{D^{(1)}_{1}e^{-iq_1x}\over\sqrt q_1})\sin{{\pi}\over a}(y+{a\over 2})
\label{eq22}
\end{equation}
\begin{equation}
\psi(III)^{(1)}_2 = ({C^{(1)}_{2}e^{iq_2x}\over\sqrt q_2}+{D^{(1)}_{2}e^{-iq_2x}\over\sqrt q_2})\sin{2{\pi}\over a}(y+{a\over 2})
\label{eq23}
\end{equation}
$A^{(1)}_{1}$, $B^{(1)}_{1}$, $A^{(1)}_{2}$, $B^{(1)}_{2}$ are the amplitudes of wave functions in
upper arm and $C^{(1)}_{1}$, $D^{(1)}_{1}$, $C^{(1)}_{2}$, 
$D^{(1)}_{2}$ are the amplitudes of wave functions in lower arm for incidence of electrons
along channel 1. Similarly, $A^{(2)}_{1}$, $B^{(2)}_{1}$, $A^{(2)}_{2}$, 
$B^{(2)}_{2}$ are the amplitudes of wave functions in upper arm  and 
$C^{(2)}_{1}$, $D^{(2)}_{1}$, $C^{(2)}_{2}$, $D^{(2)}_{2}$ are the 
amplitudes of wave functions in lower arm for incidence of electrons along channel 2.
Depending on the choice of energy $ E $ and potential $ V_0 $, $ q_1 $ 
and $ q_2 $ can be real (propagating mode) as well as imaginary (evanescent mode). 
Such evanescent states can always be excited in the internal regions of the system but not 
in leads \cite{wee,wha}.
In the regimes II and III,
we can choose $V_0$ to be non-zero such that for $n=1,2$, 
$ E < V_0 + {n^2 \pi^2 \hbar^2\over 2m^{*}a^2}$. Then we are considering both
channels (modes) to be evanescent. For such evanescent channels 
$q_n$ in Eq. (\ref{eq20}) and Eq. (\ref{eq23}) is replaced by $ is_n$ 
and wave functions can be written as
\begin{equation}
\psi(II)^{(1)}_1 = ({A^{(1)}_{1}e^{-s_1x}\over\sqrt {is_1}}+{B^{(1)}_{1}e^{s_1x}\over\sqrt {is_1}})\sin{{\pi}\over a}(y+{a\over 2}) 
\label{eq24}
\end{equation}
\begin{equation} 
\psi(II)^{(1)}_2 = ({A^{(1)}_{2}e^{-s_2x}\over\sqrt {is_2}}+{B^{(1)}_{2}e^{s_2x}\over\sqrt {is_2}})\sin{2{\pi}\over a}(y+{a\over 2}) 
\label{eq25}
\end{equation}
\begin{equation}
\psi(III)^{(1)}_1 = ({C^{(1)}_{1}e^{-s_1x}\over\sqrt {is_1}}+{D^{(1)}_{1}e^{s_1x}\over\sqrt {is_1}})\sin{{\pi}\over a}(y+{a\over 2}) 
\label{eq26}
\end{equation}
\begin{equation}
\psi(III)^{(1)}_2 = ({C^{(1)}_{2}e^{-s_2x}\over\sqrt {is_2}}+{D^{(1)}_{2}e^{s_2x}\over\sqrt {is_2}})\sin{2{\pi}\over a}(y+{a\over 2})
\label{eq27}
\end{equation}
Note that this form of the wave functions along with its imaginary normalization 
constant is necessary to give the correct expression for currents in evanescent 
modes. We can also choose the constant potential $V_0$ and the 
$n$ value of $ n^2 \pi^2 \hbar^2 \over 2m^{*}a^2 $ in such a fashion that one channel (or mode)
is propagating and the other channel (or mode) is evanescent and we can describe
the wavefunctions appropriately.

\begin{figure}[h!]
\centering
\includegraphics[height=7.truecm,width=9.truecm]{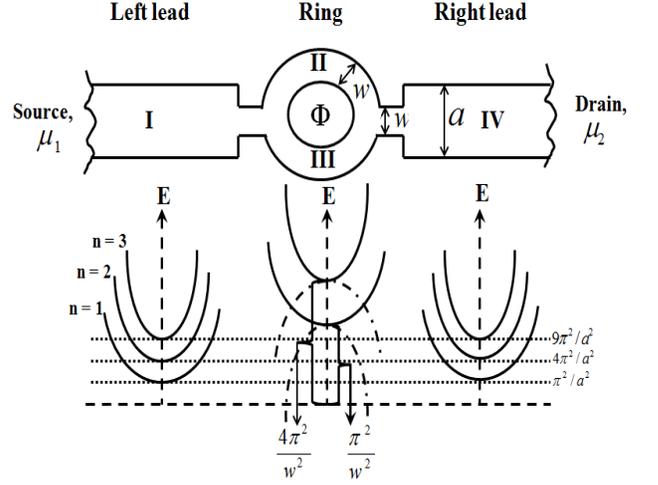}
\caption{\small Top: Schematic diagram of another mesoscopic interferometer
coupled to the left and right electron reservoirs. Electrostatic potential 
is zero everywhere.
Bottom: Energy level dispersion curve for each region. 
Dispersion relation, $ E(k_n)$ vs. $k_n$ obtained from Eq. (\ref{eq9}) 
for propagating modes in regions I and IV, indicated by solid lines.
Dispersion relation, $ E(q'_n)$ vs. $q'_n$ obtained from Eq. (\ref{eq28}) and Eq. (\ref{eq29})
for propagating modes in regions II and III, indicated by solid lines. 
Dispersion relation, $ E(s_n)$ vs. $s_n$ obtained from Eq. (\ref{eqa}) and Eq. (\ref{eqb})
for evanescent modes in regions II and III, indicated by dash-dotted lines.}
\label{fig3}
\end{figure}

In Fig. \ref{fig2} the potential $ V_0 $ in the shaded region need not be made
by an electrostatic field. One can do it by designing the system as shown 
in Fig. \ref{fig3}. 
Fig. \ref{fig3} represents the schematic diagram of a mesoscopic interferometer
(made up of normal metal or semiconductor). The width of the
quantum wire $ a $, is greater than the width of quantum ring $ w $. 
Electrostatic potential is $0$ everywhere. For the regions II and III 
in Fig. \ref{fig3}, one can obtain just like in Eq. (\ref{eq18}) and (\ref{eq19}),
\begin{eqnarray}
q'_1 & = & \sqrt{{2m^*E\over {\hbar}^2} - {{\pi}^2\over {w^2}}} \nonumber\\
& = & \sqrt{{2m^*E\over {\hbar}^2} - {{\pi}^2\over {a^2}}+{{\pi}^2\over {a^2}}-{{\pi}^2\over {w^2}}} \nonumber\\
& = & \sqrt{{2m^*(E - V'_0)\over{\hbar}^2} - {{\pi^2}\over{a^2}}}
\label{eq28}
\end{eqnarray}
where $ {{\pi}^2\over {a^2}}-{{\pi}^2\over {w^2}} = - {2m^*V'_0\over {\hbar}^2} $.
Similarly, in the second mode
\begin{equation}
q'_2 = \sqrt{{2m^*(E - V''_0)\over{\hbar}^2} - {{4\pi^2}\over{a^2}}}
\label{eq29}
\end{equation}
where $ {4{\pi}^2\over {a^2}}-{4{\pi}^2\over {w^2}} = - {2m^*V''_0\over {\hbar}^2} $.
Therefore, $V'_0$ and $V''_0$ play the same role as $V_0$ in Eq. (\ref{eq18}) and (\ref{eq19}). However,
it simply originates from geometric parameters and not from an electrostatic 
potential. The dispersion relation $E(S_n)$ for the evanescent mode
wavefunctions (Eq. (\ref{eq24}) and (\ref{eq27})) in regions II and III can also 
be obtained from Eq. (\ref{eq28}) and Eq. (\ref{eq29}) using the fact
$ q'_n \rightarrow is_{n} $. Therefore, they are given by
\begin{equation}
s_1 = \sqrt{{{\pi}^2\over {w^2}} - {2m^*E\over {\hbar}^2}}
\label{eqa}
\end{equation}
\begin{equation}
s_2 = \sqrt{{{4\pi}^2\over {w^2}} - {2m^*E\over {\hbar}^2}}
\label{eqb}
\end{equation}

The dispersion relations for different regions
(I, II and III, IV) are also shown in Fig. \ref{fig3}.  
In regions I and IV, from Eq. (\ref{eq9}) we see that E is given by 
$ E(k_n) = {\hbar^2 k^{2}_n \over 2m^*} + {n^2 \pi^2 \hbar^2 \over 2m^{*}a^2}$, 
where $ n = 1, 2, 3,... $, denotes the modes. These dispersion 
relations for different modes indexed by $n$ are indicated by solid lines (see Fig. \ref{fig3}).
The offset values are obtained at $ {\pi^2 \over a^2}$ (for $n = 1$), $ {4\pi^2 \over a^2}$ (for $n = 2$) etc and are
indicated by dotted lines. 
In regions II and III, $E(q'_n)$ is given by Eq. (\ref{eq28}) and Eq. (\ref{eq29}) 
and the offset is $ {\pi^2 \over w^2}$ and $ {4\pi^2 \over w^2}$ for
the two modes shown ($ n=1$ and $ n=2$).
Here the offset values are 
indicated by the second brackets. There are no propagating states
between the dotted lines at $ {\pi^2 \over a^2}$ (for $n = 1$) 
and $ {9\pi^2 \over a^2}$ (for $n = 3$) within the ring but there are propagating states 
in the leads. In these energy regimes, the electrons tunnel through as 
evanescent modes described by Eq. (\ref{eq24}) - Eq. (\ref{eq27}). Their
dispersion curve $E(s_n)$ as obtained from Eq. (\ref{eqa}) and Eq. (\ref{eqb})
are shown by dash-dotted lines.

\begin{figure}[h!]
\centering
\includegraphics[height=3.truecm,width=6.truecm]{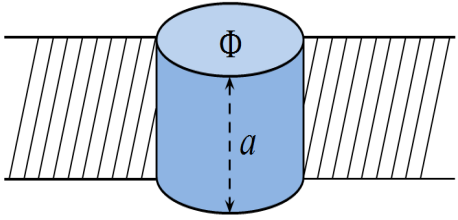}
\caption{\small Cylindrical Aharonov-Bohm set up}
\label{fig4}
\end{figure}

Note that a two dimensional quantum wire can be also converted into a Aharonov-Bohm set 
up as shown in Fig. \ref{fig4}. Essentially one can form a cylinder that can enclose a flux 
\cite{sha}. In this case all the analysis given above remains the same.
For example if now we choose cylindrical coordinates the wave function
in Eq. (\ref{eq20}) becomes 
\begin{equation}
\psi(II)^{(1)}_{1} = ({A^{(1)}_{1}e^{im_1\theta}\over\sqrt m_1}+{B^{(1)}_{1}e^{-im_1\theta}\over\sqrt m_1})\sin{{\pi}\over a}(z+{a\over 2})
\label{eq30}
\end{equation}
In fact this makes analysis much simpler because $ m_1$ stands for angular momentum and 
takes into account the curvature of the ring. The advantage of this set up
is already described by S. Mukherjee et al. \cite{sr}.

In our previous work \cite{sr} we proposed a junction scattering
matrix $S$ for a multi-channel junction that can be easily generalized
to any number of channels. One can match the wave functions depicted in 
Fig. \ref{fig2} 
at junction J1 and J2 and conserve the currents by 
using these $S$-matrices that give us a set of linear equations. 
For evanescent modes the internal wavefunctions has to be appropriately
chosen, given by Eq. (\ref{eq24}) - Eq. (\ref{eq27}). We can 
calculate the coefficients $ A^{(1)}_{1} $, $ A^{(1)}_{2} $, $ B^{(1)}_{1} $, 
$ B^{(1)}_{2} $, $ A^{(2)}_{1} $, $ A^{(2)}_{2} $, $ B^{(2)}_{1} $, $ B^{(2)}_{2} $, 
$ C^{(1)}_{1} $, $ C^{(1)}_{2} $, $ D^{(1)}_{1} $, $ D^{(1)}_{2}$, $ C^{(2)}_{1} $, 
$ C^{(2)}_{2} $, $ D^{(2)}_{1} $ and $ D^{(2)}_{2} $ by matrix inversion. \\

The general definition of current ($ I $) is given by
\begin{equation}
 I = \int_{-\frac{a}{2}}^{\frac{a}{2}} \frac{e\hbar}{2im^{*}}(\Psi^\dag\vec{\bigtriangledown}\Psi-\Psi\vec{\bigtriangledown}\Psi^\dag) dy 
\label{eq31}
\end{equation}
Suppose there are two channels (modes) in the ring, which are both propagating. Current 
($ I $) in propagating mode is obtained by calculating Eq. (\ref{eq31})
for wave functions given by equations from Eq. (\ref{eq20}) to Eq. (\ref{eq23}). For such
propagating modes in upper arm, for incidence along channel 1, the partial current is given by
\begin{equation}
 I^{(1)}_{U}(pp) = \frac{e\hbar}{m^{*}}[|A^{(1)}_{1}|^2 - |B^{(1)}_{1}|^2 + |A^{(1)}_{2}|^2 - |B^{(1)}_{2}|^2]
\label{eq32}
\end{equation}
For propagating modes in upper arm, for incidence along channel 2, the partial current is given by
\begin{equation}
 I^{(2)}_{U}(pp) = \frac{e\hbar}{m^{*}}[|A^{(2)}_{1}|^2 - |B^{(2)}_{1}|^2 + |A^{(2)}_{2}|^2 - |B^{(2)}_{2}|^2]
\label{eq33}
\end{equation}
Similarly, for propagating modes in lower arm, for incidence along channel 1 and 2, the 
partial currents are given by
\begin{equation}
 I^{(1)}_{L}(pp) = \frac{e\hbar}{m^{*}}[|C^{(1)}_{1}|^2 - |D^{(1)}_{1}|^2 + |C^{(1)}_{2}|^2 - |D^{(1)}_{2}|^2]
\label{eq34}
\end{equation}
and
\begin{equation}
 I^{(2)}_{L}(pp) = \frac{e\hbar}{m^{*}}[|C^{(2)}_{1}|^2 - |D^{(2)}_{1}|^2 + |C^{(2)}_{2}|^2 - |D^{(2)}_{2}|^2]
\label{eq35}
\end{equation}
When both modes are propagating using Eqs. (\ref{eq32}) and (\ref{eq33}) we can write the net current 
in upper arm as 
\begin{equation}
 I_{U} = I^{(1)}_{U}(pp) + I^{(2)}_{U}(pp) 
\label{eq36}
\end{equation}
When both modes are propagating using Eqs. (\ref{eq34}) and (\ref{eq35}) we can write the net current 
in lower arm as 
\begin{equation}
 I_{L} = I^{(1)}_{L}(pp) + I^{(2)}_{L}(pp) 
\label{eq37}
\end{equation}
Next we consider that there are two channels (modes)
in the ring, one being propagating and other being evanescent. Current ($ I $) 
using one propagating mode and one evanescent mode is obtained by
calculating Eq. (\ref{eq31}) choosing the wavefunctions appropriately for
propagating mode and evanescent mode given in Eq. (\ref{eq20}) to Eq. (\ref{eq27}).
In this case in upper arm for incidence along channel 1 the partial current is given by
\begin{eqnarray}
I^{(1)}_{U}(pe) 
& = & \frac{e\hbar}{m^{*}}[|A^{(1)}_{1}|^2-|B^{(1)}_{1}|^2] \nonumber\\
& + & \frac{e\hbar}{im^{*}}[{A^{(1)}_{2}}^{*}B^{(1)}_{2}e^{-i\alpha}-A^{(1)}_{2}{B^{(1)}_{2}}^{*} e^{i\alpha}]
\label{eq38}
\end{eqnarray}
For one propagating mode and one evanescent mode in upper arm, for 
incidence along channel 2, the partial current is given by
\begin{eqnarray}
I^{(2)}_{U}(pe) 
& = & \frac{e\hbar}{m^{*}}[|A^{(2)}_{1}|^2-|B^{(2)}_{1}|^2] \nonumber\\
& + & \frac{e\hbar}{im^{*}}[{A^{(2)}_{2}}^{*}B^{(2)}_{2}e^{-i\alpha}-A^{(2)}_{2}{B^{(2)}_{2}}^{*}e^{i\alpha}]
\label{eq39}
\end{eqnarray}
Similarly, for one propagating mode and one evanescent mode in lower arm, 
for incidence along channel 1 and 2, the partial currents are given by
\begin{eqnarray}
 I^{(1)}_{L}(pe) 
& = & \frac{e\hbar}{m^{*}}[|C^{(1)}_{1}|^2-|D^{(1)}_{1}|^2] \nonumber\\
& + & \frac{e\hbar}{im^{*}}[{C^{(1)}_{2}}^{*}D^{(1)}_{2}e^{-i\alpha}-C^{(1)}_{2}{D^{(1)}_{2}}^{*} e^{i\alpha}]
\label{eq40}
\end{eqnarray}
and
\begin{eqnarray}
 I^{(2)}_{L}(pe) 
& = & \frac{e\hbar}{m^{*}}[|C^{(2)}_{1}|^2-|D^{(2)}_{1}|^2] \nonumber\\
& + & \frac{e\hbar}{im^{*}}[{C^{(2)}_{2}}^{*}D^{(2)}_{2}e^{-i\beta}-C^{(2)}_{2}{D^{(2)}_{2}}^{*} e^{i\beta}]
\label{eq41}
\end{eqnarray}
For using one propagating mode and one evanescent mode the net current in upper arm is given by
\begin{equation}
 I_{U} = I^{1}_{U}(pe) + I^{2}_{U}(pe) 
\label{eq42}
\end{equation}
For using one propagating mode and one evanescent mode the net current in lower arm is given by
\begin{equation}
 I_{L} = I^{(1)}_{L}(pe) + I^{(2)}_{L}(pe) 
\label{eq43}
\end{equation}
Next we consider that both the channels are evanescent.
The current ($ I $) for two evanescent modes is obtained by
calculating Eq. (\ref{eq31}) choosing the wavefunctions appropriately 
for evanescent modes given in Eq. (\ref{eq24}) to Eq. (\ref{eq27}).
For such evanescent modes the partial current in upper arm, for incidence along channel 1 is given by
\begin{eqnarray}
 I^{(1)}_{U}(ee) 
& = & \frac{e\hbar}{im^{*}}[{A^{(1)}_{1}}^{*}B^{(1)}_{1}e^{-i\alpha}-A^{(1)}_{1}{B^{(1)}_{1}}^{*}e^{i\alpha} \nonumber\\
& + & {A^{(1)}_{2}}^{*}B^{(1)}_{2}e^{-i\alpha}-A^{(1)}_{2}{B^{(1)}_{2}}^{*}e^{i\alpha}]
\label{eq44}
\end{eqnarray}
For evanescent mode the partial current in upper arm, for incidence along channel 2 is given by
\begin{eqnarray}
I^{(2)}_{U}(ee) & = & \frac{e\hbar}{im^{*}}[{A^{(2)}_{1}}^{*}B^{(2)}_{1}e^{-i\alpha}-A^{(2)}_{1}{B^{(2)}_{1}}^{*}e^{i\alpha} \nonumber\\
& + & {A^{(2)}_{2}}^{*}B^{(2)}_{2}e^{-i\alpha}-A^{(2)}_{2}{B^{(2)}_{2}}^{*}e^{i\alpha}]
\label{eq45}
\end{eqnarray}
Similarly, for evanescent mode in lower arm and incidence along 
channel 1 and 2, the partial currents are given by
\begin{eqnarray}
 I^{(1)}_{L}(ee)
& = & \frac{e\hbar}{im^{*}}[C^{(1)}_{1}{D^{(1)}_{1}}^{*}e^{i\beta}-{C^{(1)}_{1}}^{*}D^{(1)}_{1}e^{-i\beta}  \nonumber\\ 
& + & C^{(1)}_{2}{D^{(1)}_{2}}^{*}e^{i\beta}-{C^{(1)}_{2}}^{*}D^{(1)}_{2}e^{-i\beta}]
\label{eq46}
\end{eqnarray}
and
\begin{eqnarray}
 I^{(2)}_{L}(ee) 
& = & \frac{e\hbar}{im^{*}}[C^{(2)}_{1}{D^{(2)}_{1}}^{*}e^{i\beta}-{C^{(2)}_{1}}^{*}D^{(2)}_{1}e^{-i\beta} \nonumber\\  
& + & C^{(2)}_{2}{D^{(2)}_{2}}^{*}e^{i\beta}-{C^{(2)}_{2}}^{*}D^{(2)}_{2}e^{-i\beta}] 
\label{eq47}
\end{eqnarray}
From Eqs. (\ref{eq44}) and (\ref{eq45}) we can write for evanescent modes the net 
current in upper arm is given by
\begin{eqnarray}
 I_{U} = I^{(1)}_{U}(ee) + I^{(2)}_{U}(ee)
\label{eq48}
\end{eqnarray}
Similarly, from Eqs. (\ref{eq46}) and (\ref{eq47}) we can write for evanescent modes 
net current in lower arm is given by
\begin{equation}
 I_{L} = I^{(1)}_{L}(ee) + I^{(2)}_{L}(ee)
\label{eq49}
\end{equation}

\section{ Results and Discussions}
\subsection{\it Conductance of a multi-channel A-B ring}
We are considering a two channel A-B ring that is 
characterized by four transmission amplitudes $ t'^{(1)}_{1} $ , $ t'^{(1)}_{2} $,  $ t'^{(2)}_{1} $ and 
$ t'^{(2)}_{2} $ and four reflection amplitudes $ r'^{(1)}_{1} $, $ r'^{(1)}_{2} $, $ r'^{(2)}_{1} $ and 
$ r'^{(2)}_{2} $.
Landauer's formula gives the two probe conductance $ G $ as
\begin{equation}
G= \frac{2e^2}{h}\sum_{i,j}|t'^{(i)}_{j}|^2.
\label{eq50}
\end{equation}
The transmission amplitude from mode $ i $ to mode $ j $ is $ t'^{(i)}_{j} $. $ G $ 
is a strongly oscillating function of $ \phi/\phi_0 $ implying we can use 
flux to drive the system from a conducting state to an insulating state. 
In case of a triode, we can use grid voltage to change the 
current flow from cathode to anode and therefore use it as a switch 
or a transistor. Similarly, in this A-B set up we can use 
magnetic field to control the current from source to drain and similarly 
we can use it as a switch or a transistor. 

\begin{figure}[h]
\begin{center}
\includegraphics[height=7.truecm,width=8.truecm]{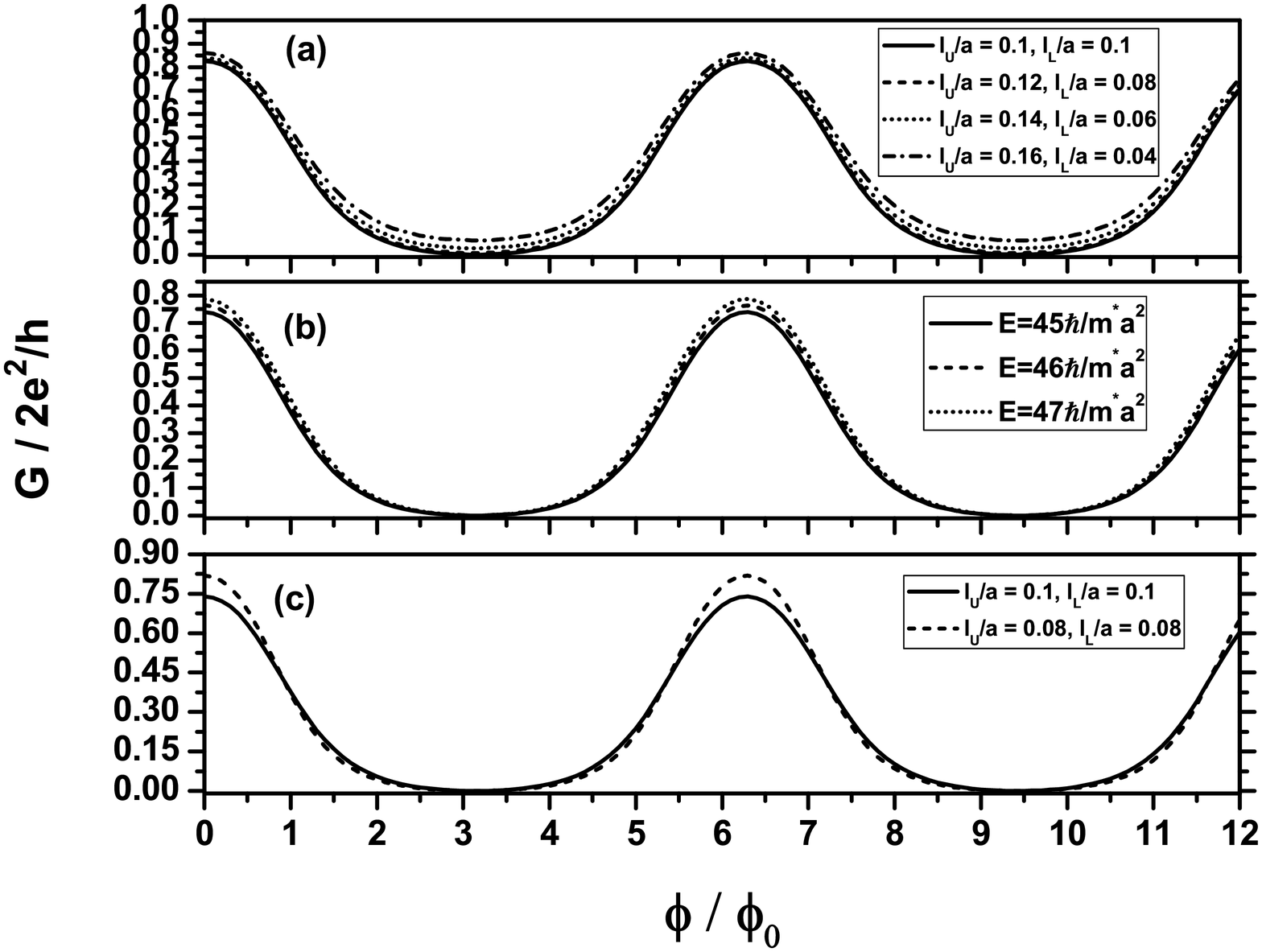}
\caption{\footnotesize{The figures (a) - (c) show plots of conductance, $G/{2e^2\over h}$
as a function of $ \phi/\phi_0 $ using evanescent modes for different 
arm lengths (a), for different Fermi energy (b) and for different ring
lengths (c). Here $ \phi_0 = hc/e $. 
The exact values of $ Em^*a^2/\hbar^2 $, $l_U$ and $l_L$ are given in the figure inset. For all sets of 
different parameters $G/{2e^2\over h}$ as a function of $ \phi/\phi_0 $
have the same nature.}}
\label{fig5}
\end{center}
\end{figure}

In an A-B ring, switching action is based on constructive and 
destructive interference and is extremely sensitive to small 
changes in parameters like Fermi energy, ring length, arm lengths 
etc. They are practically impossible to control \cite{land}. This
fact changes completely if we use evanescent modes \cite{sr}. 
An electron in an evanescent mode do not acquire phase changes 
associated with propagation or impurity scattering. Only phase 
changes are due to A-B effect and we find that within a period 
(0 to $ 2\pi $) conductance is maximum (or minimum) at zero flux, 
then it goes through a deep minimum (or a maximum) and rises 
(or falls) again to a maximum (or minimum) value. One can 
explain this as follows. Conductance being a symmetric 
function of flux (Onsager reciprocality relation), is a 
function of $ (\cos n\phi/\phi_0) $. So it maximizes (or minimizes) 
at $ 0 $ flux and then decreases (or increases) with flux. 
Periodicity is always $ \phi_0 $ in absence of other competing source of phase 
changes and absence of resonance. This behavior 
is independent of all parameters. Since evanescent modes are 
not very conducting we have to take smaller rings. A plot 
of the conductance is shown in Fig. \ref{fig5} that exhibits this. 

\subsection{\it Circulating currents in a multi-channel A-B ring}

Let us consider the system in absence of flux. When $ l_U = l_L $, 
the currents $I_{S}$ through the system (source to drain currents) split at junction J1
and divides into two equal parts. The upper arm current, $I_U$ (i.e., $I_{S}/2$) 
flows in clockwise direction along the upper arm of the ring and 
the lower arm current, $I_L$ (i.e., $I_{S}/2$) flows in anti-clockwise 
direction along the lower arm of the ring. The clockwise and anti-clockwise currents are equal 
in magnitude and hence the system has no magnetization. When $ l_U \ne l_L $, 
in absence of flux, these two currents are
different in magnitude and there exists two distinct 
possibilities depending on the choice of 
$l_U$, $l_L$ ($l_U \ne l_L$) and Fermi energy. In the first
possibility, for a certain range of incident Fermi energies (or wave
vectors), the currents in the two arms $I_{U}$ and $I_{L}$ are individually
less than $I_{S}$.
Then currents in both arms flow in the direction of the
chemical potential gradient, upper arm current flows in clockwise direction and lower
arm current flows in anti-clockwise direction. In such a situation there is no circulating
currents flowing in the ring. However, in a certain energy interval,
it turns out that current in one arm (upper or lower) is larger than 
the total currents $I_{S}$ (magnification property). 
This implies that, to conserve the total currents at the junctions, 
the current in the other arm (lower or upper) should flow against the 
chemical potential gradient
In such a situation the current flow in this arm of the loop continues to flow in the 
ring as a circulating currents. 
Based on the constraints on the local currents, the existence of
circulating currents in a one-dimensional mesoscopic ring with
asymmetric arm lengths was established 
in earlier works \cite{jay1,jay2,jay3}. In 2010, Su et. al. \cite{su}
formulated the constraint condition for circulating currents as follows:
\begin{equation}
 I_c = {1\over 2}sign[I_{U}](|I_{U}| + |I_{L}| - I_{S})
\label{eq51}
\end{equation}

In presence of flux, the circulating currents $ I_{c}$ changes drastically in magnitude but the 
definition remains the same. Once we have formalized the current flowing in the
multichannel quasi one-dimensional ring we can use the same definition
to calculate the circulating currents $I_{c}$ in the system shown 
in Fig. \ref{fig2}. We now plot circulating currents as well as transmission coefficient as a 
function of incident Fermi energy for this system. 

\begin{figure}[h]
\begin{center}
\includegraphics[height=6.truecm,width=8.truecm]{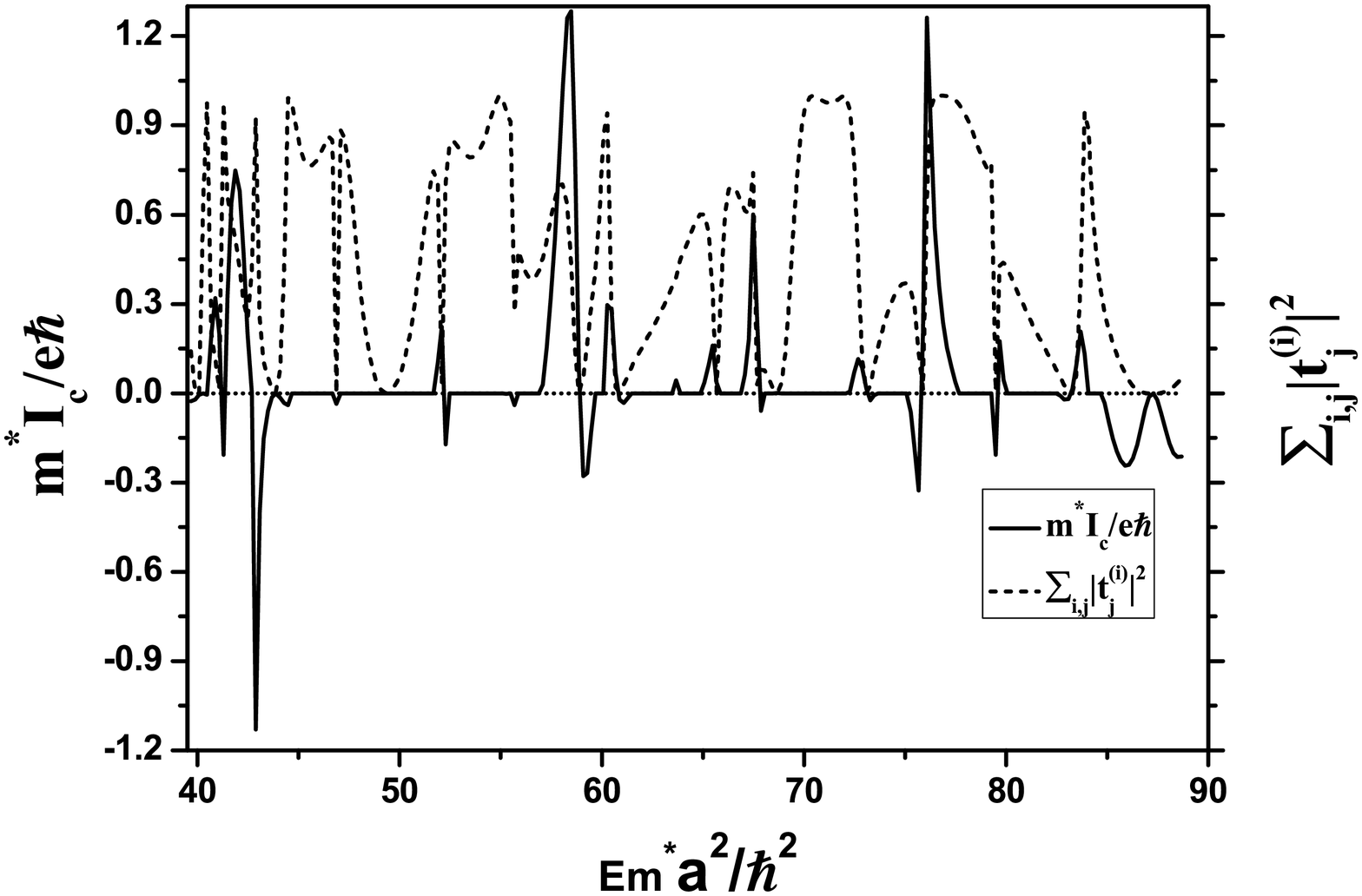}
\caption{\footnotesize{The figure shows plot of circulating currents 
($I_c$) in dimensionless unit (solid curve) and transmission coefficient ($\sum_{i,j}|t'^{(i)}_{j}|^2$) 
(dashed curve) as a function of incident Fermi energy using two 
propagating modes. Here upper arm length, $l_U/a = 3$ and lower arm length, $l_L/a= 7$.}}
\label{fig6}
\end{center}
\end{figure}

{\it Two propagating modes}:
We first consider a case when both the channels (modes) in the 
ring are propagating. 
In Fig. \ref{fig6} we have plotted the circulating currents 
$I_c$ (solid curve) as a function of incident Fermi
energy. We have taken $l_U/a = 3$ and $l_L/a= 7$. In Fig. \ref{fig6} we have also plotted
the transmission coefficient $\sum_{i,j}|t'^{(i)}_{j}|^2$ (dashed curve). 
We notice that the circulating currents
appears as we cross the energy where $\sum_{i,j}|t'^{(i)}_{j}|^2=0$.
In absence of channel mixing $\sum_{i,j}|t'^{(i)}_{j}|^2$ can have minima 
but does not become $0$. Thus inclusion of evanescent mode is very
important to study current magnification.

\begin{figure}[h]
\begin{center}
\includegraphics[height=6.truecm,width=8.truecm]{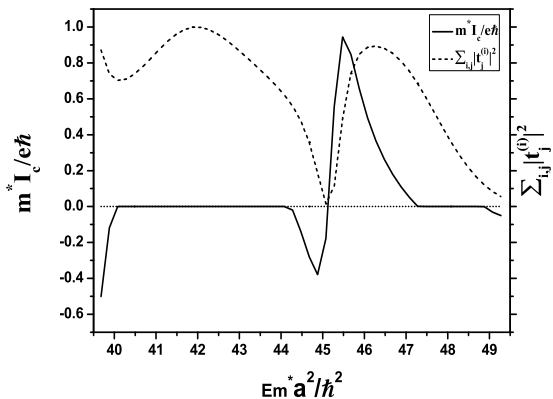}
\caption{\footnotesize{The figure shows plot of circulating 
currents ($I_c$) in dimensionless unit (solid curve) and transmission 
coefficient ($\sum_{i,j}|t'^{(i)}_{j}|^2$) (dashed curve) as a function 
of Fermi energy of incident electrons using one propagating mode and one evanescent mode. 
Here upper arm length, $l_U/a = 3$ and lower arm length, $l_L/a= 7$.}}
\label{fig7}
\end{center}
\end{figure}

{\it One propagating mode and one evanescent mode}:
Now we will consider the case where one mode is propagating
and the other mode is evanescent.
In Fig. \ref{fig7} we have plotted 
circulating currents $I_c$ (solid curve) as a function of incident 
Fermi energy. We have taken $l_U/a = 3$ and $l_L/a= 7$. 
In Fig. \ref{fig7} we have also plotted the transmission coefficient, 
$\sum_{i,j}|t'^{(i)}_{j}|^2$ (dashed curve). We again notice that the 
circulating currents appear as we cross the energy where $\sum_{i,j}|t'^{(i)}_{j}|^2=0$.

When both modes are evanescent then there is no circulating currents in
the system. This is essentially because evanescent modes are decaying modes
and for such modes $I_{U}$ or $I_{L}$ cannot be greater than $I_{S}$
for any energy.

\subsection{\it Magnetization of a multi-channel A-B ring}

The geometry in consideration (Fig. \ref{fig2}, Fig. \ref{fig3}, 
Fig. \ref{fig4}) can have a magnetization 
due to the internal currents in regions II and III. A proper formulation for
evaluating such internal currents in the regime of evanescent modes is
still not well established \cite{jay1,jay2,jay3} and the correct formalism 
that gives consistent results in every situation is given in Eqs. (\ref{eq32}) - (\ref{eq49}).
This internal currents can induce a magnetic field that can be
measured. 

\begin{figure}[h]
\begin{center}
\includegraphics[height=6.truecm,width=8.truecm]{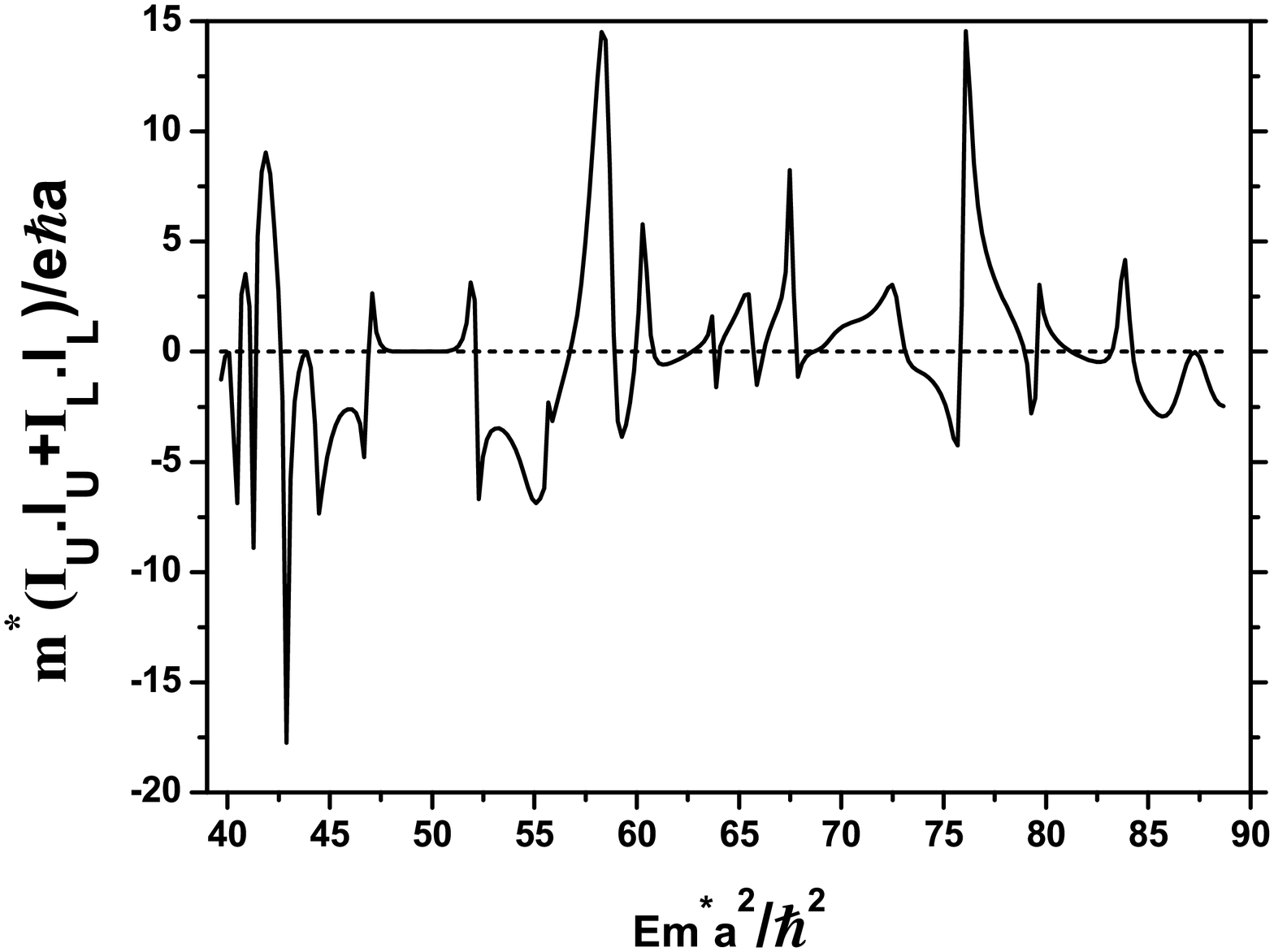}
\caption{\footnotesize{The figure shows plot of magnetization strength in 
dimensionless unit as a function of Fermi energy using two
propagating modes. Here upper arm length, $l_U/a = 3$ and lower arm length, $l_L/a= 7$.}}
\label{mgE}
\end{center}
\end{figure}

There are two different origins to this magnetization. 
First is due to transport currents and this magnetization can be 
there even in absence of the A-B flux. The currents through the sample, $ I_{S} $ 
splits into $I_U$ and $I_L$ in upper arm and lower arm, respectively. 
For $ l_U \neq l_L $, $I_U \neq I_L$ and thus the clockwise 
current is different from anticlockwise current. This can result
in a magnetization, magnetization strength is given by $ (I_{U}.l_{U} + I_{L}.l_{L}) $.
The current in lower arm (upper arm), $I_{L}$ ($I_{U}$) by definition is 
negative (positive). In regimes of circulating 
currents $I_L$ ($I_{U}$) becomes positive (negative) leading to a large enhancement in magnetization
which is shown in Fig. \ref{mgE} where we have plotted magnetization strength as 
a function of Fermi energy. Flux value is taken to be 0, and there are two propagating
modes. 
The second is due to the A-B flux and this magnetization can exist even if $ \mu_1 = \mu_2 $ and 
there is no incident current. A-B flux induces an equilibrium 
current and this current is always an odd function of flux 
\cite{pin}. The first kind of magnetization is
further modified by the flux but is always an even function 
of flux. It is known that the odd component can be determined
from the even component \cite{akk} and vice versa. 
The total magnetization is therefore a linear 
combination of an even and an odd function of flux and hence
can be any arbitrary function of flux. It will be a function of
$ \cos (n\phi/\phi_0) $ as well as $ \sin (n\phi/\phi_0) $. Therefore
unlike the conductance it will not maximize (or minimize) at 0 flux and then 
decreases (or increases) with flux. It can maximize at any arbitrary flux.
Therefore it remains to be seen if magnetization due to evanescent 
modes remain independent of material parameters. If it does then
one can build stable devices that rely on magnetization.
In practice it is not always possible to maintain either the ring length (or arm length) 
or the incident energy ($ Em^*a^2/\hbar^2 $) values constant
due to statistical fluctuation or temperature
changes.

\begin{figure}[h!]
\begin{center}
\includegraphics[height=6.truecm,width=8.truecm]{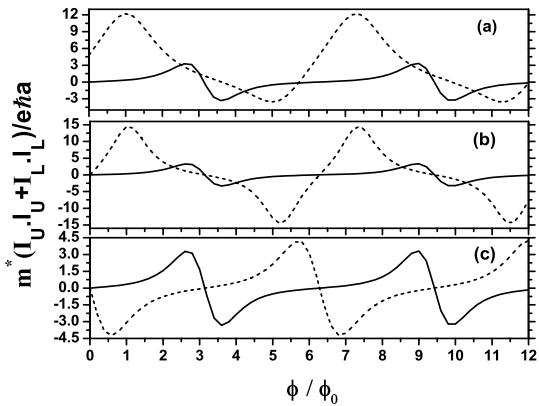}
\caption{\footnotesize{The figures (a) - (c) show plot of
magnetization strength in dimensionless unit for the system in Fig. \ref{fig2} as a function of 
$ \phi/\phi_0 $ for different choice of parameters. Here $ \phi_0 = hc/e $. 
The constant potential $ V_0$ is taken as 0, so that both the modes are propagating.
In (a), the incoming electrons have energy $ E = {55{\hbar}^2\over m^*a^2} $. 
The solid line is for $ l_{U}/a = 5, l_{L}/a = 5 $ and the dashed line is 
for $ l_{U}/a = 4, l_{L}/a = 6 $. 
In (b) the arm length ratio is taken as $l_{U}:l_{L} = 5:5$.
The solid line is for $ E = {55{\hbar}^2\over m^*a^2} $ and the dashed line is
for $ E = {57{\hbar}^2\over m^*a^2} $. 
In (c), the incoming electrons have energy $ E = {55{\hbar}^2\over m^*a^2} $. 
Keeping the arm length ratio same ($l_{U}:l_{L}=1:1$) we have taken here different ring lengths.
The solid line is for $ L/a = 10$ and the dashed line is for $L/a = 8$. }}
\label{fig8}
\end{center}
\end{figure}

{\it Two propagating modes}: Here we have taken three different cases.
In all three cases we have plotted the strength of magnetization, i.e., 
$(I_{U}.l_{U} + I_{L}.l_{L})$ as a function of $\phi/\phi_0$ for different parameters.
In the first case we have taken two different arm length values keeping the Fermi energy
and total ring length same (see Fig. \ref{fig8} (a)). In the second case we have taken two 
different Fermi energy values keeping the ratio of arm length 
and total ring length same (see Fig. \ref{fig8} (b)). In the third case we have taken two 
different ring length values keeping the Fermi energy
and ratio of arm length same (see Fig. \ref{fig8} (c)). The parameter values are
described in detail in figure caption. In the first case we can see 
that the maximum of the solid line and the dashed line
are not obtained at the same flux values. In the second case the solid line has no distinct features whereas
the dashed line has sharp maximum and minimum. In the third case the maximum and
minimum of the solid line are not in the same phase as that of the dashed line. Thus
we can see here that
the strength  of magnetization $(I_{U}.l_{U} + I_{L}.l_{L})$ as a 
function of $ \phi/\phi_0 $ is strongly dependent on 
variations in parameters like different ratio of arm lengths,
different Fermi energies, different total ring lengths etc.
Therefore we can conclude that using propagating modes we can 
not build reliable devices based on magnetic properties.

\begin{figure}[h!]
\begin{center}
\includegraphics[height=6.truecm,width=8.truecm]{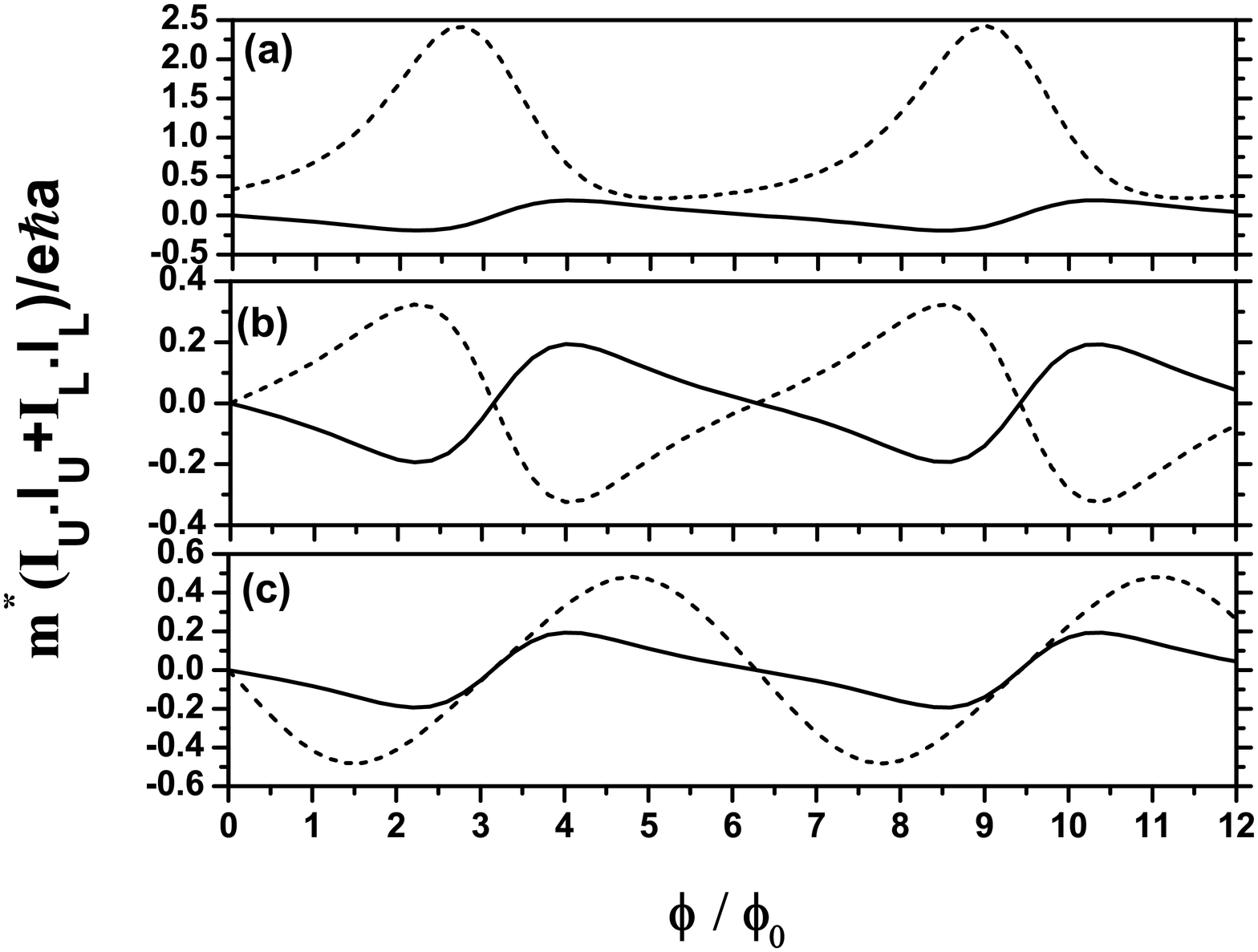}
\caption{\footnotesize{The figures (a) - (c) show plot of
magnetization strength in dimensionless unit for the system in Fig. \ref{fig2} as a function of $ \phi/\phi_0 $. The constant
potential $ V_0 $ of the ring is such that $ V_0 = 10{\hbar}^2/em^*a^2 $.
With this choice $ q_1 $ is real and $ q_2 $ is imaginary. Thus we 
are considering in this case one propagating mode and one 
evanescent mode. 
In (a), the incoming electrons have energy $ E = {45{\hbar}^2\over m^*a^2} $. 
The solid line is for $ l_{U}/a = 1, l_{L}/a = 1 $ and the dashed line is 
for $ l_{U}/a = 1.2, l_{L}/a = 0.8 $. 
In (b) the arm length ratio is taken as $l_{U}:l_{L} = 1:1$.
The solid line is for $ E = {45{\hbar}^2\over m^*a^2} $ and the dashed line is
for $ E = {47{\hbar}^2\over m^*a^2} $. 
In (c), the incoming electrons have energy $ E = {45{\hbar}^2\over m^*a^2} $. 
Keeping the arm length ratio same ($l_{U}:l_{L}=1:1$) we have taken here different ring lengths.
The solid line is for $ L/a = 2$ and the dashed line is for $L/a = 1.6$.}}
\label{fig9}
\end{center}
\end{figure}

{\it One propagating mode and one evanescent mode}:
Here we have taken three different cases.
In all three cases we have plotted the strength of magnetization, i.e., 
$(I_{U}.l_{U} + I_{L}.l_{L})$ as a function of $\phi/\phi_0$ for different parameters.
In the first case we have taken two different arm length values keeping the Fermi energy
and total ring length same (see Fig. \ref{fig9} (a)). In the second case we have taken two 
different Fermi energy values keeping the ratio of arm length 
and total ring length same (see Fig. \ref{fig9} (b)). In the third case we have taken two 
different ring length values keeping the Fermi energy
and ratio of arm length same (see Fig. \ref{fig9} (c)). The parameter values are
described in detail in figure caption. In the first case we can see 
that the solid line has no distinct features whereas
the dashed line has sharp maximum and minimum. In the second case we can see that
where the solid line has minimum the dashed line has maximum and vice versa. In the third case the maximum and
the minimum of the solid line are not in the same phase as that of the dashed line. Thus
we can see here that
the strength  of magnetization $(I_{U}.l_{U} + I_{L}.l_{L})$ as a 
function of $ \phi/\phi_0 $ is strongly dependent upon 
the variations in parameters like different ratio of arm lengths,
different Fermi energy, different total ring length etc. Therefore we can conclude that 
using one propagating mode and one evanescent mode we can 
not build reliable devices based on magnetic properties.

This random behavior of magnetization is essentially due to extreme
sensitivity of quantum interference to small changes in phase of wavefunction.
Fig. \ref{fig8} (a), (b), (c) and Fig. \ref{fig9} (a), (b), (c) exemplifies this
for some representative choice of parameters and can be easily checked for
any other choice of parameters.

\begin{figure}[h!]
\begin{center}
\includegraphics[height=6.truecm,width=8.truecm]{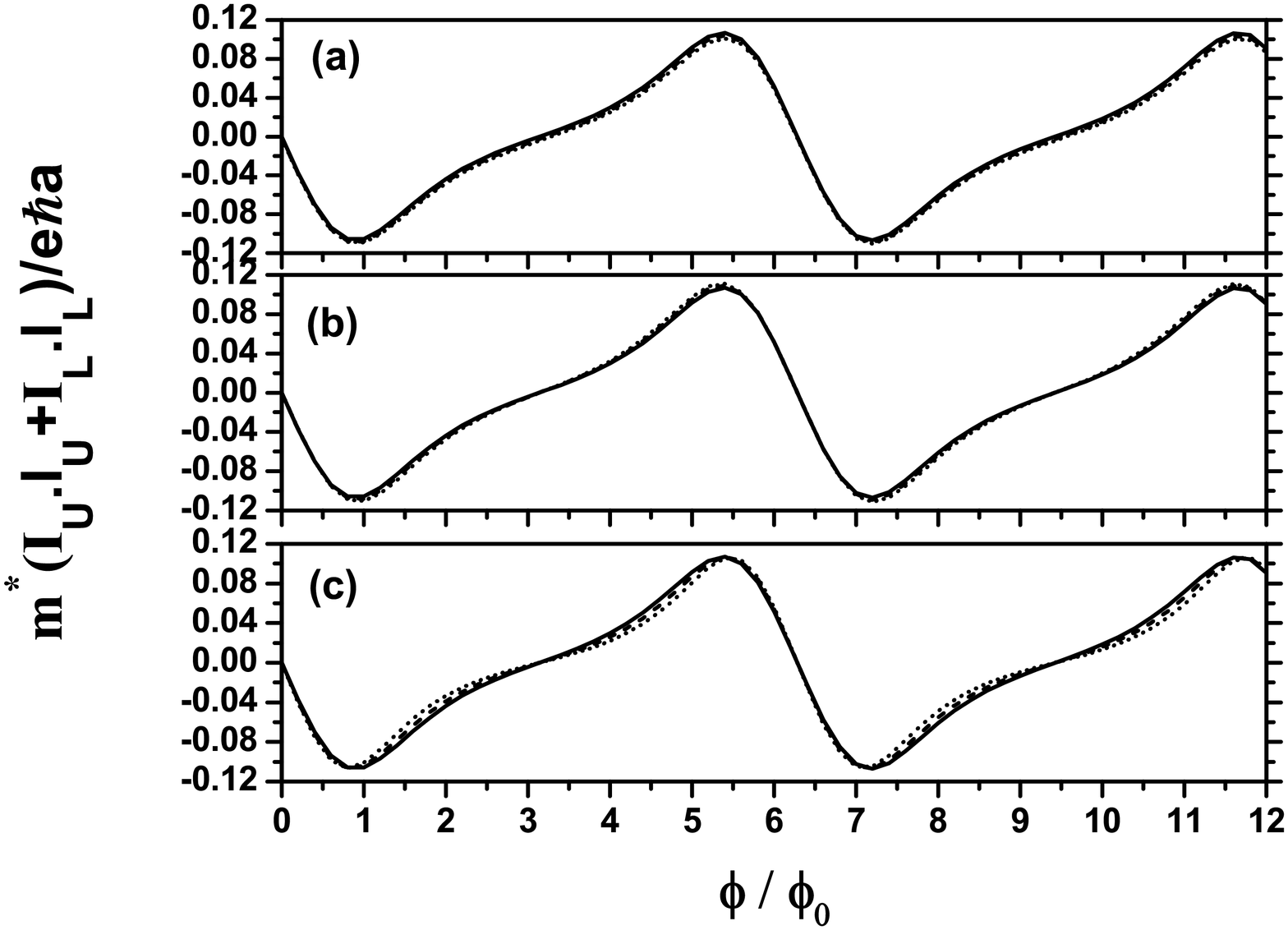}
\caption{\footnotesize{The figures (a) - (c) show plots of
magnetization strength in dimensionless unit for the system in Fig. \ref{fig2} as a function of $ \phi/\phi_0 $ for an
Aharonov-Bohm interferometer where all the current carrying modes
are evanescent. Here $ \phi_0 = hc/e $. 
In (a), the incoming electrons have energy $ E = {45{\hbar}^2\over m^*a^2} $. 
The solid line is for $ l_{U}/a = 0.1, l_{L}/a = 0.1 $, the dashed line is 
for $ l_{U}/a = 0.12, l_{L}/a = 0.08 $ and the dotted line is for $l_{U}/a = 0.14, 
l_{L}/a = 0.06 $. All three plots show nature of magnetization is independent 
of relative ratio of arm lengths. 
In (b), the arm length ratio is taken as $l_{U}:l_{L} = 0.1:0.1$.
The solid line is for $ E = {45{\hbar}^2\over m^*a^2} $, the dashed 
line is for $ E = {46{\hbar}^2\over m^*a^2} $ and the dotted line is
for $ E = {47{\hbar}^2\over m^*a^2} $. All three plots show nature 
of magnetization is independent of Fermi energy. 
In (c), the incoming electrons have energy $ E = {45{\hbar}^2\over m^*a^2} $. 
Keeping the arm length ratio same ($l_{U}:l_{L}=1:1$) we have taken here 
different ring lengths. The solid line is for $ L/a = 0.2$, the dashed line is for  $ L/a = 0.18$ 
and the dotted line is for $L/a = 0.16$. All three plots show nature of 
magnetization is independent of total ring length.}}
\label{fig10}
\end{center}
\end{figure}

{\it Two evanescent modes}: 
Here also we have taken three different cases.
In all three cases we have plotted the strength of magnetization, i.e., 
$(I_{U}.l_{U} + I_{L}.l_{L})$ as a function of $\phi/\phi_0$ for different parameters.
In the first case we have taken two different arm length values keeping the Fermi energy
and total ring length same (see Fig. \ref{fig10} (a)). In the second case we have taken two 
different Fermi energy values keeping the ratio of arm length 
and total ring length same (see Fig. \ref{fig10} (b)). In the third case we have taken two 
different ring length values keeping the Fermi energy
and ratio of arm length same (see Fig. \ref{fig10} (c)). The parameter values are
described in detail in figure caption. In all three cases we have shown that
the strength  of magnetization $(I_{U}.l_{U} + I_{L}.l_{L})$ as a 
function of $ \phi/\phi_0 $ is qualitatively as well as quantitatively same 
for different choice of parameters like different ratio of arm lengths,
different Fermi energies, different total ring lengths etc.
So there is possibility of robust device action based
on magnetic response using only evanescent modes which contradicts
earlier works that devices based on quantum interference effects
cannot be achieved. Note that the maximum or the minimum in figures 
(Fig. \ref{fig10} (a), (b) and (c)) is not at 0 flux as we have argued. 
However the maxima and minima are fixed and do not change with change 
in parameters. The reason for this is explained later.

So far we have considered three different cases - (i) keeping Fermi 
energy value and ring length fixed, we have chosen three diffrent
sets of arm lengths, (ii) keeping arm length and ring length fixed, 
we have chosen three diffrent sets of Fermi energies, and (iii) keeping Fermi 
energy value and arm length fixed, we have chosen three diffrent
sets of ring lengths. In Fig. 11 we will discuss another set of plots where we 
have chosen incident Fermi energy and electrostatic potential
inside the ring in such a fashion that both the channels are
evanescent and sample parameters vary randomly. Sample parameter
choice is described in the figure caption. Here too we show
that magnetic behavior for random choice of parameters is  qualitatively as well as quantitatively 
the same.

\begin{figure}[h!]
\begin{center}
\includegraphics[height=6.truecm,width=8.truecm]{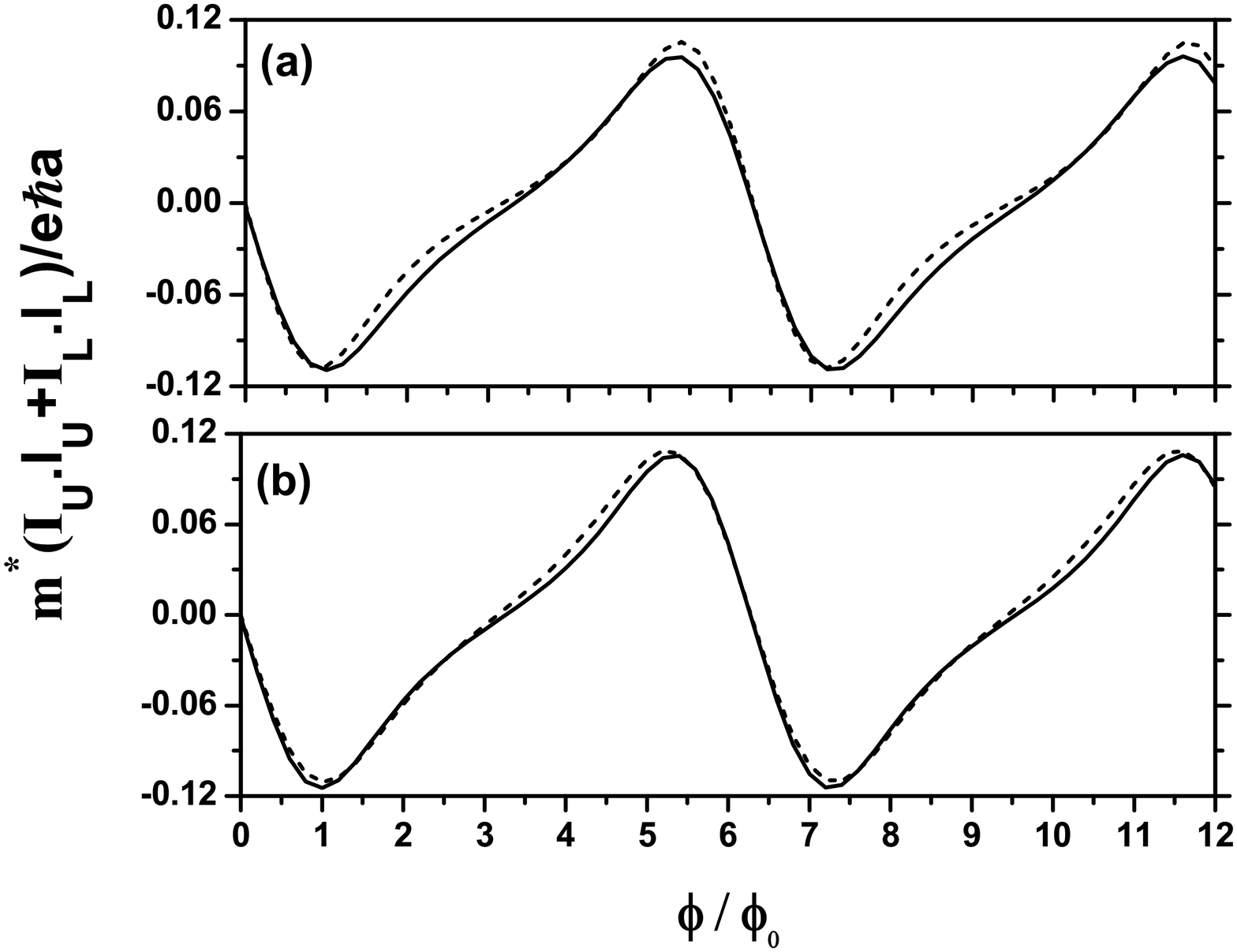}
\caption{\footnotesize{The figures (a) - (b) show plot of
magnetization strength for the system in Fig. \ref{fig2} as a function of $ \phi/\phi_0 $ for an
Aharonov-Bohm interferometer where all the current carrying modes
are evanescent. Here $ \phi_0 = hc/e $. 
In (a), the incoming electrons have energy $ E = {45{\hbar}^2\over m^*a^2} $.
The solid line is for $L/a = 0.24$ and the dashed line is for
$l_{U}/a = 0.11, l_{L}/a = 0.09$. 
In (b), the incoming electrons have energy $ E = {47{\hbar}^2\over m^*a^2} $.
The solid line is for $L/a = 0.22$ and the dashed line is for
$l_{U}/a = 0.13, l_{L}/a = 0.12$.
Here (a) and (b) shows that plots are independent of material parameters.}}
\label{fig11}
\end{center}
\end{figure}

\begin{figure}[h!]
\begin{center}
\includegraphics[height=6.truecm,width=8.truecm]{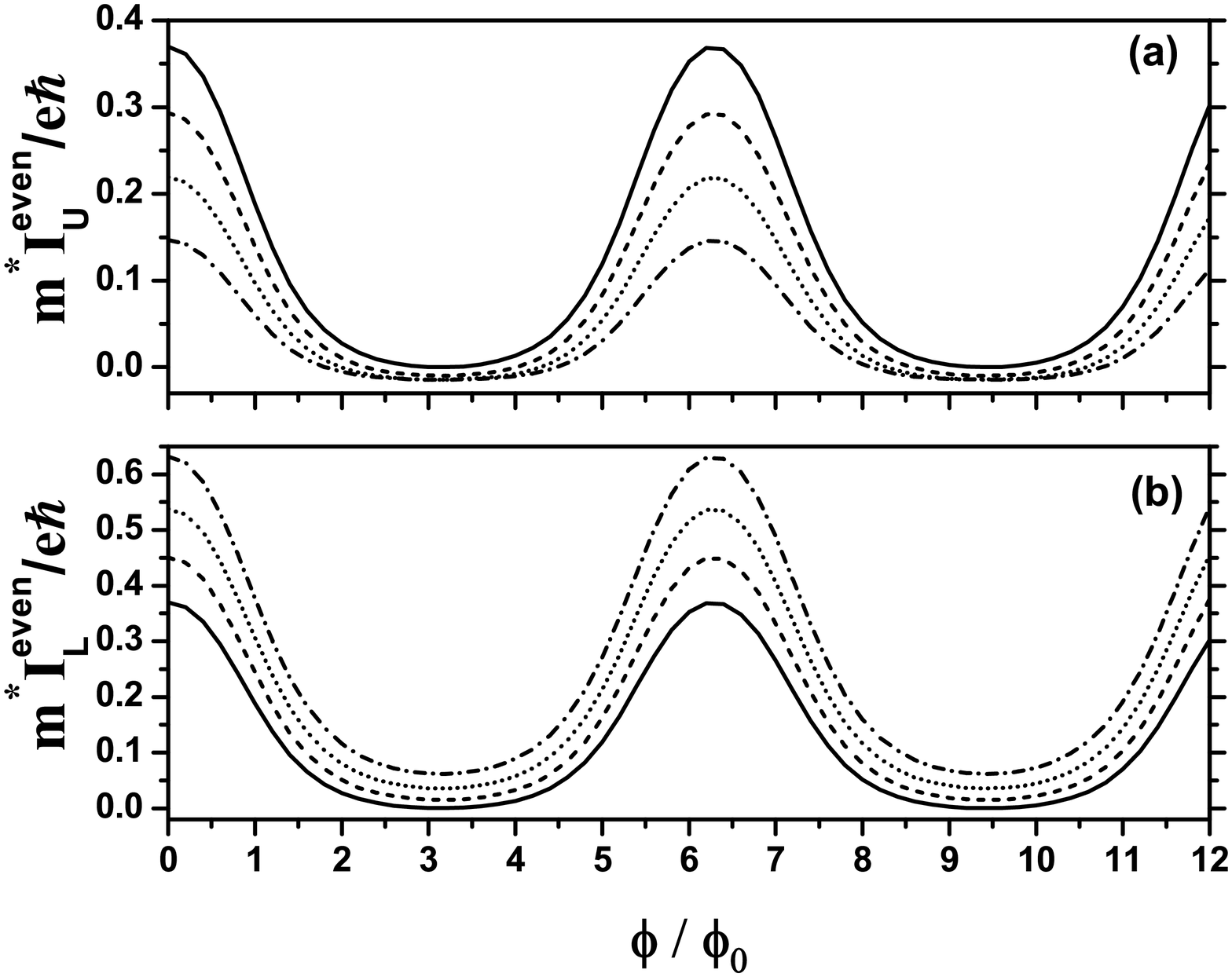}
\caption{\footnotesize{The figures (a) - (b) show plots of transport 
currents $I^{even}_{U/L}$ in dimensionless unit for the system in Fig. \ref{fig2} as a function of $ \phi/\phi_0 $ for
different arm lengths keeping the ring length same. 
In (a) and (b), the solid line is for arm length values of $l_U/a=0.1, l_L/a = 0.1$, 
the dashed line is for arm length values of $l_U/a=0.12, l_L/a = 0.08$, the dotted 
line is for arm length values of $l_U/a=0.14, l_L/a = 0.06$, and the dash-dotted line is
for arm length values of $l_U/a=0.16, l_L/a = 0.04$.}}
\label{fig12}
\end{center}
\end{figure}

As we have argued before the current in the upper arm ($I_{U}$) 
and that in the lower arm ($I_{L}$) has an even (in $ \phi $) 
contribution and an odd contribution. This uniform behavior of
magnetization can be understood by analyzing the even and odd
components separately. The even current is
transport current which is given by 
$$ I^{even}_{U,L} = {I_{U,L}(\phi) + I_{U,L}(-\phi) \over 2} $$
and plotted in Fig. \ref{fig12}. In Fig. \ref{fig12} (a), we have plotted the
transport current of the upper arm ($ I^{even}_{U}$)
as a function of flux and in Fig. \ref{fig12} (b), we have plotted transport current of the 
lower arm ($ I^{even}_{L} $) as a function of flux. Keeping the total ring 
length same we have taken four different choices of arm length ratio and in all four
cases we have shown that the nature of the transport currents (in terms of
position of maximum and minimum)
in both upper arm and lower arm remains the same that is 
independent of arm length ratio. In both these figures peak value,
 i.e., $ I^{max} $ is obtained at $ \phi/\phi_0 = 6.2 $. In Fig. \ref{fig12} (a), with the increasing
value of the upper arm length the $ I^{max}_{U} $ value decreases and in 
Fig. \ref{fig12} (b), with the decreasing value of the lower arm length the
$ I^{max}_{L} $ value increases.

\begin{figure}[h!]
\begin{center}
\includegraphics[height=6.truecm,width=8.truecm]{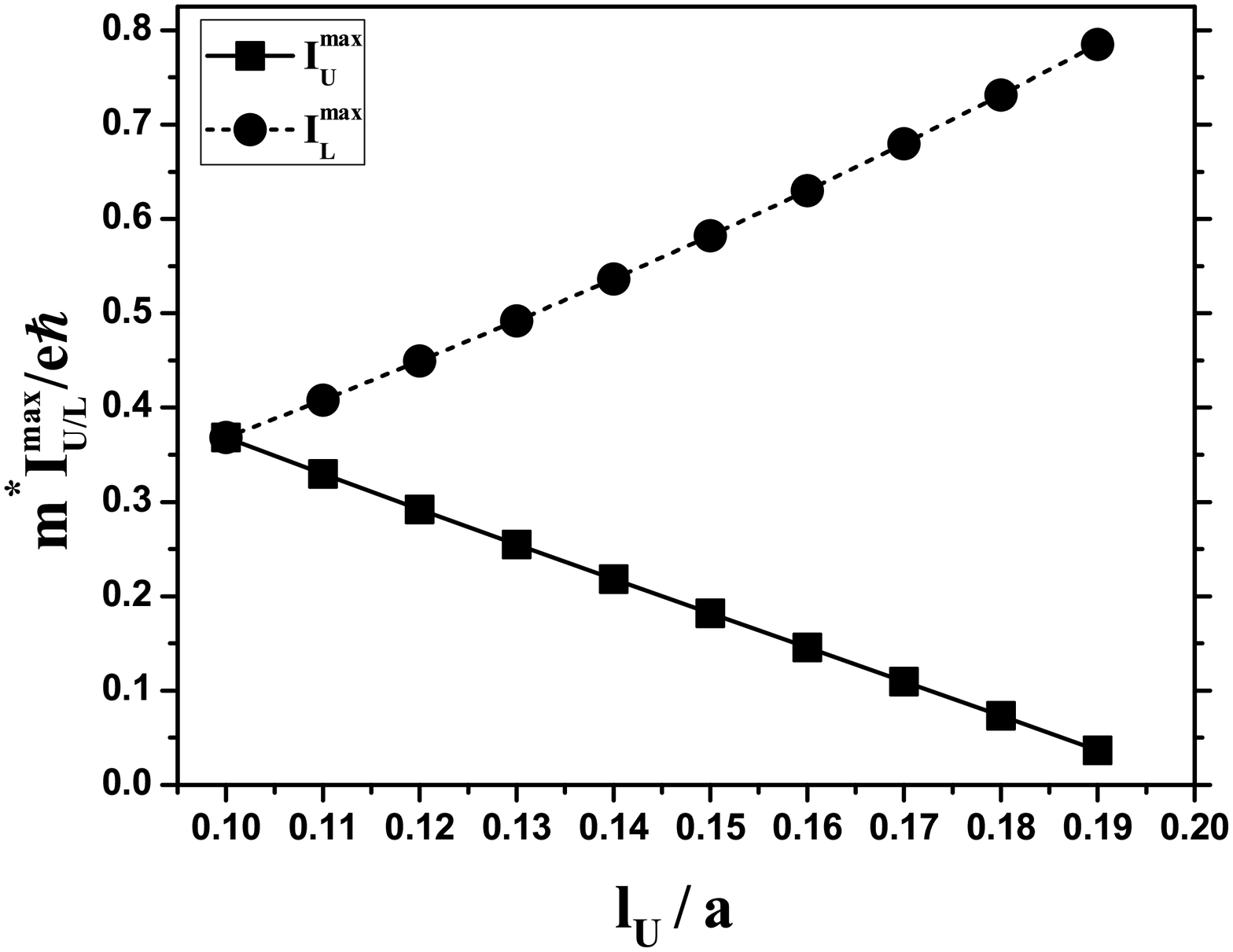}
\caption{\footnotesize{The figure shows plot of $m^{*}I^{max}_{U/L}/e\hbar$
obtained at $ \phi/\phi_0 = 6.2 $ (see Fig. 12)
as a function of upper arm length ($l_{U}/a$) of the quantum ring.
The solid line indicates $m^{*}I^{max}_{U}/e\hbar$ as a function of $l_{U}/a$
and the dashed line indicates $m^{*}I^{max}_{L}/e\hbar$ as a function of 
$l_{U}/a$.}}
\label{fig13}
\end{center}
\end{figure}

One would have thought that since the evanescent modes decay 
exponentially inside the barrier, $ I^{max}_{U/L} $ would 
have scaled exponentially with the length. But counterintuitively 
they scale linearly. The peak value ($I^{max}_{U/L}$), obtained 
at $ \phi/\phi_0 = 6.2 $, is plotted along the arm length in Fig. \ref{fig13}. 
The solid line indicates the peak values for the upper arm 
($I^{max}_{U}$) for different arm length and the dashed line
indicates the peak values for the lower arm ($I^{max}_{L}$). 
With increasing the upper arm length, $I^{max}_{U}$ linearly 
decreases whereas $I^{max}_{L}$ linearly increases. This
phenomenon has been reported earlier for evanescent modes 
in one dimension \cite{jay2}. We find it to occur even in presence of 
multiple modes with mixing between the modes.
Thus the upper arm and the lower arm behave as classical 
Ohmic conductors. This is the reason why magnetization curves 
are so uniform and peak magnetization is always at $ \phi/\phi_0 = 6.2 $ 
in Fig. \ref{fig12}. However, absolute quantum behavior can be seen from 
the fact that the $ I_{U}.l_{U} + I_{L}.l_{L} \neq 0 $. For a 
classical Ohmic resistor doubling of length would have halved the current
that helps in defining a material specific resistivity. Such a 
resistivity cannot be defined for evanescent modes.

\begin{figure}[h!]
\begin{center}
\includegraphics[height=6.truecm,width=8.truecm]{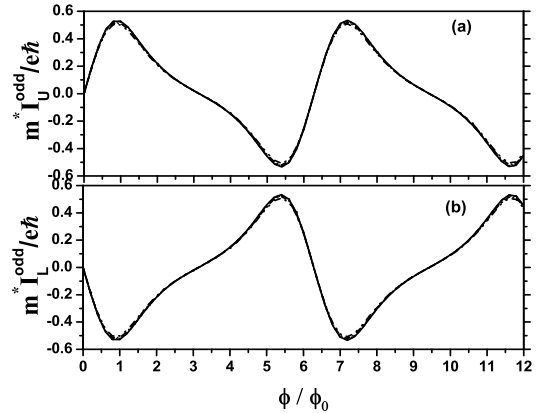}
\caption{\footnotesize{The figures (a) - (b) show plot of
persistent currents in dimensionless unit for the system in Fig. \ref{fig2} as a function of $ \phi/\phi_0 $ 
for different arm lengths keeping the ring length same. 
In (a) and (b), we have used the same convention and same parameters
as in Fig. \ref{fig12} (a) and (b), respectively.}}
\label{fig14}
\end{center}
\end{figure}

The odd current is persistent current which is given by
$$ I^{odd}_{U,L} = I_{U,L}(\phi) - I^{even}_{U,L} $$
and plotted as a function of $ \phi/\phi_0 $ in Fig. \ref{fig14}
for different arm length ratio described in figure caption. 
The Fig. \ref{fig14} (a), shows the plot of persistent currents as 
a function of $ \phi/\phi_0 $ for the upper arm and 
Fig. \ref{fig14} (b) shows the same for the lower arm. 
The nature of Fig. \ref{fig14} (a) and (b) remains the 
same when we change the arm length ratio. The flux value at 
which the curves peak also remain the same.
This is the reason why the total magnetization which is a linear
combination of the odd component and even component also remain the
same in nature and magnitude as sample parameters and Fermi
energy is varied randomly.

\section{Conclusions}
Although one-dimensional quantum rings coupled to reservoirs have received
a lot of attention in the past, realistic multi-channel rings were never
considered. We have considered such a quantum ring coupled to two
reservoirs. We have developed the correct formalism to include 
evanescent modes and multi-channel mixing. Interesting 
quantum phenomena arise in such systems like persistent
currents, circulating currents, Aharonov-Bohm effect in conductance or transport currents. 
We have studied all these quantum phenomena for a realistic 
multi-channel ring. Earlier stable quantum switch device was
proposed based on evanescent mode conductance and Aharonov-Bohm
interferometry \cite{jay1,sr} (see Fig. \ref{fig5} and Fig. \ref{fig12}).
Internal currents never received any attention and can have equal
importance in device application.
We have calculated the
circulating currents for two different cases. First we have considered two
propagating modes along the quantum ring and then we have considered one 
propagating mode and one evanescent mode along the quantum ring. We
cannot get circulating currents using evanescent modes only.
We have also calculated magnetization of the quantum ring. Magnetization 
in quantum ring arises due to two reasons. First is due 
to transport currents and second is due to Aharonov-Bohm effect. We have 
calculated strength of magnetization for three different cases, 
(i) taking two propagating modes, (ii) taking one propagating mode and 
one evanescent mode and (iii) taking two evanescent modes. For the first
two cases we have shown that magnetization behavior is very sensitive for variations in 
parameters like ring length, arm length, Fermi energy etc, while 
that for evanescent modes the strength  of magnetization 
($I_{U}.l_{U} + I_{L}.l_{L}$) as a 
function of $ \phi/\phi_0 $ is same for all variations in system parameters. 
Hence there is possibility of robust device action based
on magnetic response using evanescent modes only (see Fig. \ref{fig10}, 
Fig. \ref{fig11} and Fig. \ref{fig12}). 

\section*{Acknowledgment}
SM and PSD acknowledges Institute of Physics, Bhubaneswar for 
providing local hospitality where a part of this research work 
has been done and we all thank DST, India for financial support.


\end{document}